\documentclass[9pt,twocolumn,twoside]{osajnl}
\journal{ol}
\setboolean{shortarticle}{true}
\usepackage{graphicx}
\usepackage{txfonts}
\usepackage[english]{babel}
\usepackage{xr}
\newlength{\figsize}
\newlength{\subfigsize}
\makeatletter
\newcommand{\manuallabel}[2]{\def\@currentlabel{#2}\label{#1}}
\makeatother
\manuallabel{TR_Csca_lam_GaP_lattrecthole_bz1.38}{S2}
\manuallabel{TR_Csca_lam_GaP_lattrecthole_da25}{S5}
 
\begin{document}

\title{Enhanced Optical Kerr Effect in Metasurfaces Featuring Arrays of Rotated Rectangular Holes via Trapped-Mode Resonances}


\author{Andrey V. Panov}


 \affil[1]{
Institute of Automation and Control Processes,
Far Eastern Branch of Russian Academy of Sciences,
5, Radio st., Vladivostok, 690041, Russia}
\affil[*]{Corresponding author: panov@iacp.dvo.ru}
\dates{Received 4 June 2025; revised 21 July 2025; accepted 19 August 2025; published 11 September 2025}
\setboolean{displaycopyright}{true}
\setcounter{page}{5849}
\renewcommand*{\journalname}{Vol. 50, No. 18 / 15 September 2025 / Optics Letters}
\doi{\url{https://doi.org/10.1364/OL.569837}}

\begin{abstract}
\normalsize
Recent advances in nanophotonics have demonstrated that various optical resonances in nanostructures can achieve strong field confinement with substantially suppressed scattering. 
This study investigates the optical Kerr effect (OKE) enhancement in high-refractive-index metasurfaces featuring non-BIC trapped-mode resonances, using gallium phosphide (GaP) as a model material. 
Three-dimensional finite-difference time-domain simulations reveal a significant enhancement of the effective second-order refractive index, reaching values up to 700 times greater than bulk GaP.
The numerical results show strong electromagnetic field localization at the trapped-mode resonance, characterized by a wide transmittance dip (low Q-factor) and near-unity reflectance. 
Remarkable stability is observed, with the OKE enhancement maintaining less than 5\% variation for moderate polarization angles and tolerating random nanohole rotations that model fabrication imperfections.
Compared to bound states in the continuum (BIC), the non-BIC trapped mode demonstrates superior robustness to structural disorder while achieving comparable nonlinear enhancement. 
These findings suggest promising applications in reflective nonlinear optics, particularly for high harmonic generation and polarization-insensitive photonic devices.
\end{abstract}


\maketitle


Metasurfaces, artificial two-dimensional materials engineered to control electromagnetic waves with subwavelength precision, have garnered considerable attention due to their potential in various optical applications.
Among the various resonant phenomena supported by metasurfaces, trapped modes, also known as dark states, are particularly intriguing. 
In metasurfaces, a trapped mode refers to a resonant electromagnetic mode that is highly confined within the structure and exhibits minimal coupling to the far field. 
The dark states are typically associated with resonances that are confined due to the geometry or material properties of the structure.
The trapped modes can occur in various types of waveguides, cavities, and periodic structures like metasurfaces. 
These modes can be engineered by designing structures with specific symmetries (e.g., broken symmetry), allowing certain modes to be decoupled from free-space radiation and effectively ``trapped'' within the structure.
The trapped modes, being a theoretical abstraction, in practice are frequently referred to as quasi-trapped modes, which have weak coupling with radiation and moderate quality (Q) factors.
The trapped modes are closely related to bound states in the continuum (BICs), a special class of trapped modes that exist within the radiation continuum yet remain perfectly confined.

Currently, BICs are attracting considerable attention from researchers. 
The more general class of non-BIC trapped (dark) states in the metasurfaces remains less explored.

The trapped modes are characterized by strong field enhancement, making them useful for applications such as sensing \cite{Roberts17,Zografopoulos21,Deng21}, energy harvesting \cite{Nguyen20}, lasing \cite{Droulias17,Droulias18,Prokhorov23,Prokhorov24}, nonlinear optics \cite{Marinica07,Shabanov09,Tong16}, designing narrowband filters, modulators, and other optical devices.
At present, case studies on the nonlinear optical properties of trapped modes are primarily limited to higher harmonic generation.
In the theoretical studies \cite{Marinica07,Shabanov09}, the second harmonic intensity generated at the trapped state in periodic arrays of thin dielectric cylinders was five orders of magnitude greater than that in the off-resonance case.
A silicon metasurface consisting of symmetric spindle-shape nanoparticle array exhibited 300 times larger third harmonic generation than that of bulk silicon slab \cite{Tong16}.

Previously, numerical simulations demonstrated the possibility of an anapole state with high transmission and strong electromagnetic energy confinement in high-index metasurfaces featuring an array of square holes \cite{Panov23}. 
The anapole state in metasurfaces is a non-radiating electromagnetic mode characterized by a destructive interference between electric dipole and toroidal dipole moments, leading to suppressed far-field scattering but strong near-field confinement.
Near the anapole state, the absolute value of the effective nonlinear Kerr coefficient of the metasurface was shown to be up to three orders of magnitude greater than that of an unstructured film.

In the present study, the optical Kerr effect (OKE), also known as the intensity-dependent refractive index, for the metasurface possessing trapped state is investigated with three-dimensional finite-difference time-domain (FDTD) simulations.
The existence of the trapped state arises from symmetry break of square hole lattice \cite{Panov23} by changing the hole shape to rectangle and rotating it.


The current work utilizes a numerical method for determining the effective Kerr nonlinearity of nanocomposites, based on the approach introduced in Ref.~\cite{Panov18}. This procedure involves performing three-dimensional finite-difference time-domain (FDTD) simulations of a Gaussian beam propagating through a sample exhibiting optical nonlinearity. By calculating the phase shift of the transmitted beam at various light intensities $I$, it is possible to evaluate the nonlinear refractive index that arises due to OKE as defined by
$ n=n_0+n_2 I$,
where $n_0$ is the linear refractive index and $n_2$ is the second-order nonlinear refractive index. 
The technique enables the estimation of the nonlinear refractive index at multiple points within the transmitted beam, allowing for the calculation of both the mean value and standard deviation of $n_2$.
This study is focused on the effective nonlinear refractive index of the metasurface, representing the ensemble-averaged optical properties of the sample.

The FDTD simulations, which are carried out using the MIT Electromagnetic Equation Propagation (MEEP) FDTD solver \cite{OskooiRo10}, model the Gaussian beam propagation through the nonlinear metasurface. For visible light simulations, the FDTD computational domain measures $2.8\times 2.8\times 15$~$\mu$m, with a spatial resolution of 3 nm, refined to 2.3 nm near resonances. 


The effective second-order nonlinear refractive index of the metasurface is then compared with that of bulk GaP, with all values represented as unitless quantities. The simulations are conducted for a wavelength of $\lambda = 532$~nm (unless otherwise specified in the context) and gallium phosphide (GaP) is chosen as the material due to its high refractive index ($n_{0\,\mathrm{in}} = 3.49$) at this wavelength and negligible extinction coefficient \cite{Aspnes83}.

\begin{figure}
{\centering\includegraphics[width=\figsize]{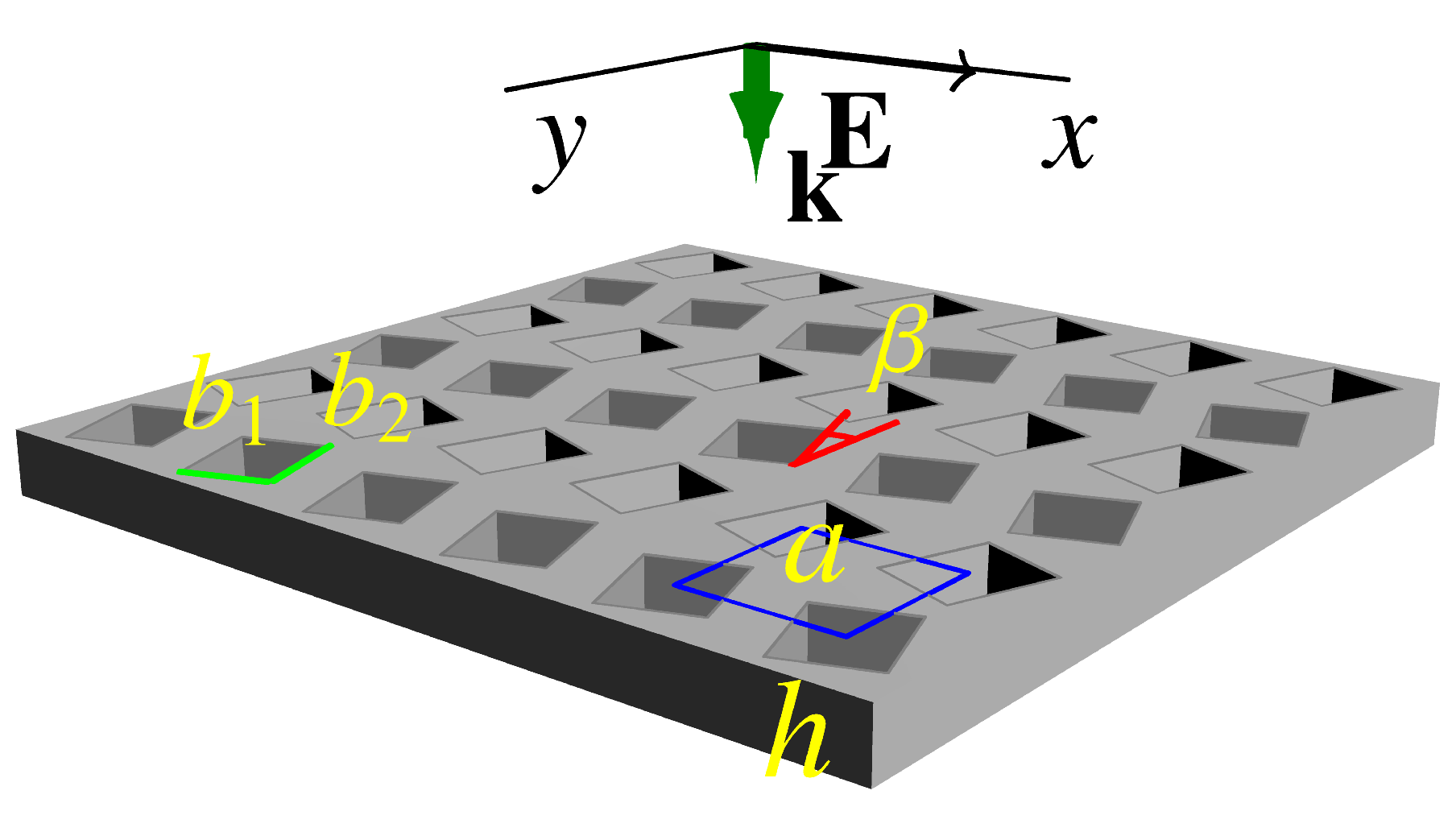}\par} 
\caption{\label{rect_doubang_pores}  
Schematic of the simulated metasurface, consisting of a square lattice array of rectangular nanoholes in a high refractive index slab. 
The holes (with side lengths  $b_1$ and $b_2$) are rotated by an angle $\beta$.
The Gaussian beam is incident normally on the metasurface. }
\end{figure}

In this research, symmetry breaking in the lattice of square holes is achieved by distorting the holes into rectangular shapes and rotating them.
The rotation is applied row-wise: alternate rows are rotated clockwise and counterclockwise around the geometric centers of the holes by an angle $\beta$.
Figure~\ref{rect_doubang_pores} shows a schematic representation of the metasurface with a lattice of rotated rectangular holes. 
The rectangular holes have side lengths $b_1$ and $b_2$, while $a$ denotes the square lattice parameter.
Optimal values for the GaP slab thickness ($h=100$~nm) and ($a=250$~nm) were obtained previously \cite{Panov23}.

\begin{figure}[tb!]
{\centering
\begin{tabbing}
\includegraphics[height=\subfigsize]{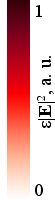} \= 
\includegraphics[width=\subfigsize]{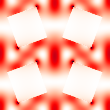}\=
\hspace{1em}\=
\includegraphics[width=\subfigsize]{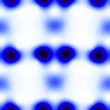}\= 
\includegraphics[height=\subfigsize]{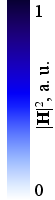}\\
 \> \includegraphics[width=\subfigsize]{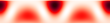} \>
  \>
 \includegraphics[width=\subfigsize]{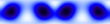}
\end{tabbing}
\par}
\caption{\label{ener_dist_GaP_144_lattrecthol} 
Time-averaged distributions of the electric ($\varepsilon|\mathbf{E}|^2$; left panel, red) and magnetic ($|\mathbf{H}|^2$; right panel, blue) energy densities in a GaP slab with rotated rectangular nanoholes at the trapped-mode resonance. Parameters: lattice constant $a = 250$~nm, slab thickness $h = 100$~nm, nanohole dimensions $b_1 = 140$~nm and $b_2 = 148$~nm, rotation angle $\beta = 20^\circ$, and wavelength $\lambda = 532$~nm. Cross-sectional views show the distributions at $z = h/2$ in the $y$-$x$ plane (upper row, under vertically polarized Gaussian beam illumination) and at $x=0$ in the $y$-$z$ plane (lower row).}
\end{figure}

First, the electric and magnetic energy distributions of the distorted square lattices of the rectangular nanoholes are calculated using the FDTD method.
The effect of symmetry breaking is investigated by varying the ratio $b_1/b_2$ and the rotation angle $\beta$. After the FDTD simulations, it is found that the electric and magnetic energy have maximum values when $b_1=140$~nm, $b_2=148$~nm, and $\beta=20^\circ$.
Figure~\ref{ener_dist_GaP_144_lattrecthol} illustrates the time-averaged electric ($\varepsilon|\mathbf{E}|^2$) and magnetic ($|\mathbf{H}|^2$) energy distributions at the cross-section of the metasurface ($h/2$).
The vector distribution of $\mathbf{E}$ in this cross-section is provided in Fig.~\ref{efield_dist_GaP_144_lattrecthol} (Supplement~1).

\begin{figure}[tb!]
{\centering
\includegraphics[width=\figsize]{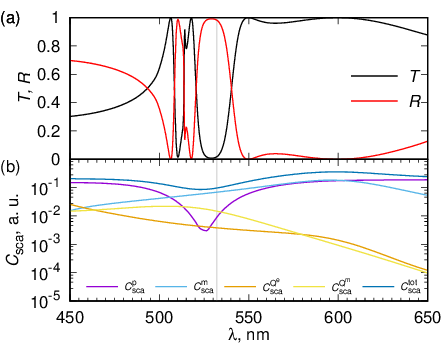}
\par} 
\caption{\label{TRspectr_GaP_rotated}
(a) Transmittance ($T$) and reflectance ($R$) spectra for the array of the rotated rectangular nanoholes. 
(b) Scattering cross-section spectra showing multipole contributions: electric dipole ($C_{\mathrm{sca}}^{\mathrm{p}}$), magnetic dipole ($C_{\mathrm{sca}}^{\mathrm{m}}$), electric quadrupole ($C_{\mathrm{sca}}^{\mathrm{Q^e}}$), and magnetic quadrupole ($C_{\mathrm{sca}}^{\mathrm{Q^m}}$), along with their sum ($C_{\mathrm{sca}}^{\mathrm{tot}}$). 
Results are shown for the rotated nanohole lattice element in the GaP slab (outlined in blue in Fig.~\ref{rect_doubang_pores})
Geometric parameters for both graphs: $a = 250$~nm, $h = 100$~nm, $b_1 = 140$~nm, $b_2 = 148$~nm, and $\beta = 20^\circ$. 
Refractive index $n_{0\,\mathrm{in}}$ is assumed to be constant over the whole wavelength range.}
\end{figure} 

The transmittance and reflectance spectra of the array of the rotated rectangular nanoholes in the GaP slab with $a=250$~nm, $h=100$~nm, $b_1=140$~nm, $b_2=148$~nm, and $\beta=20^\circ$ are depicted in Fig.~\ref{TRspectr_GaP_rotated}(a). 
There is a wide dip in the transmittance at $\lambda=532$~nm, while the narrower resonances are shifted to shorter wavelengths.
The transmittance at this $\lambda$ is close to zero, and the reflectance is close to unity.
This is a typical feature of trapped states.
In contrast, the bound states in the continuum exhibit very narrow dips in the transmittance.
The multipole decomposition for the lattice element delineated by the blue square in Fig.~\ref{rect_doubang_pores} is illustrated by 
Fig.~\ref{TRspectr_GaP_rotated}(b).
Figures \ref{TR_Csca_lam_GaP_lattrecthole_bz1.38}--\ref{TR_Csca_lam_GaP_lattrecthole_da25}  (Supplement~1) display the transmittance and reflectance spectra and the multipole decomposition results for different values of $b_2$ or $\beta$.
As can be seen, there are no scattering peaks at the wavelength of interest.
Simultaneously, this structure exhibits the electromagnetic energy confinement, confirming the existence of a generic (non-BIC) trapped (dark) state in the metasurface.
The confinement of electromagnetic energy is necessary for enhancing nonlinear optical effects.


\begin{figure}[tbh!]
{\centering
\includegraphics[width=\figsize]{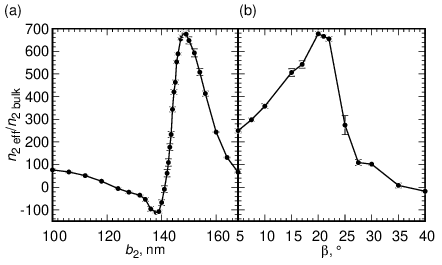}
\par} 
\caption{\label{n2_hl_GaP_lattrecthol}
Enhancement of the effective second-order refractive index $n_{2{\mathrm{eff}}}/n_{2{\mathrm{bulk}}}$ for a rotated rectangular nanohole array in a GaP slab as functions of the nanohole side length $b_2$ (a) and rotation angle $\beta = 20^\circ$ (b). Geometric parameters: $a = 250$~nm, $h = 100$~nm, $b_1 = 140$~nm, $\beta = 20^\circ$ (a) or $b_2 = 148$~nm (b).}
\end{figure} 

%

In what follows, the nonlinear optical properties (effective OKE) of the metasurface are investigated. 
Figure~\ref{n2_hl_GaP_lattrecthol} shows the dependencies of the enhancement of the effective second-order refractive index of the rotated rectangular nanohole array in GaP slab, $n_{2{\mathrm{eff}}}/n_{2{\mathrm{bulk}}}$, on the elongation of the rectangles and the rotation angle.
These dependencies exhibit maximum values of $n_{2{\mathrm{eff}}}$ about 700 times larger than the unpatterned GaP, which can be attributed to the strong field localization and enhancement within the nanohole array structure.
Similar magnitudes of the $n_{2{\mathrm{eff}}}/n_{2{\mathrm{bulk}}}$ were calculated for other types of GaP metasurfaces \cite{Panov23,Panov24}, suggesting that this enhancement factor is generic for dielectric metasurfaces with properly designed resonant elements.
The dependence of $n_{2{\mathrm{eff}}}/n_{2{\mathrm{bulk}}}$ on length of the rectangular hole $b_2$ displays a dip and a peak.
The negative values of the effective $n_{2{\mathrm{eff}}}$ near resonances in all-dielectric metasurfaces have been previously observed in numerical modeling \cite{Panov23,Panov24,Panov19} and experiments \cite{Shi25}.
A similar behavior in metal-dielectric nanocomposites is described, within the framework of effective medium theory, as interference between the local and incident fields inside the nanoparticle (meta-atom) \cite{Kohlgraf-Owens08,Jia15}.
Although the size of all-dielectric meta-atoms is much larger, the explanation may be similar.

\begin{figure}[tbh!]
{\centering
\includegraphics[width=\figsize]{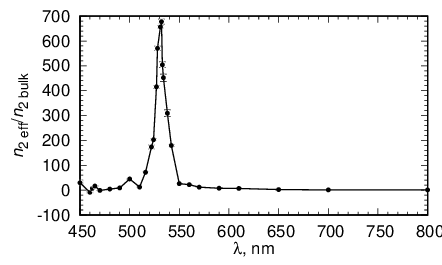}
\par} 
\caption{\label{n2_lam_GaP_144_lattrecthol_a2.5_h1}
Enhancement of the effective second-order refractive index for a rotated rectangular nanohole array in a GaP slab with parameters $a = 250$~nm, $h = 100$~nm, $b_1 = 140$~nm, and $b_2 = 148$~nm, shown as a function of wavelength. 
The refractive index $n_{0\,\mathrm{in}}$ varies according to Ref.~\cite{Aspnes83}.}
\end{figure} 

Figure~%
\ref{n2_lam_GaP_144_lattrecthol_a2.5_h1}
depicts the spectral dependence of the enhancement of the effective OKE for the rotated rectangular nanohole array in the GaP slab with $a=250$~nm, $h=100$~nm, $b_1=140$~nm, and $b_2=148$~nm.
Here, the linear refractive index $n_{0\,\mathrm{in}}$ of GaP varies with $\lambda$ according to Ref.~\cite{Aspnes83}.
This spectral dependence shows a sharp peak near $532$~nm, corresponding to the observed trapped state.
Interestingly, this peak is significantly narrower than the transmittance deep in Fig.~\ref{TRspectr_GaP_rotated}, although the base of the $n_{2{\mathrm{eff}}}$ peak coincides with this dip.
Notably, the sharp resonances of the transmittance spectra between $500$--$520$~nm do not correspond to large magnitudes of the effective OKE.

\begin{figure}[tb!]
{\centering
\includegraphics[width=\figsize]{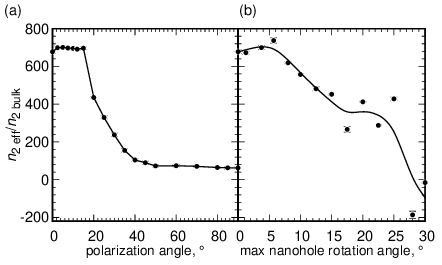}
\par} 
\caption{\label{n2_yrot_GaP_144_lattrecthol}
Enhancement of the effective second-order refractive index for a rotated rectangular nanohole array in a GaP slab with $a = 250$~nm, $h = 100$~nm, $b_1 = 140$~nm, $b_2 = 148$~nm at $\lambda = 532$~nm as functions (a) of the polarization angle (the angle between the Gaussian beam polarization direction and $x$-axis) or (b) of the maximum random rotation angle (the additional angle added to $\beta$ for each nanohole). Here the random rotation angle has continuous uniform distribution limited by the maximum random rotation angle in both directions.}
\end{figure} 


The dependence of effective OKE enhancement $n_{2{\mathrm{eff}}}/n_{2{\mathrm{bulk}}}$ on a Gaussian beam polarization angle (defined as the angle between the electric field vector $\mathbf{E}$ and the $x$-axis in Fig.~\ref{rect_doubang_pores}) is shown in Fig.~\ref{n2_yrot_GaP_144_lattrecthol}(a).
The enhancement ratio remains remarkably stable for polarization angles up to $17^\circ$ with variations of less than 5\% observed within this angular range. 
Beyond this threshold angle, $n_{2{\mathrm{eff}}}/n_{2{\mathrm{bulk}}}$ undergoes a rapid decrease, eventually saturating at a value of approximately 60 for angles exceeding $60^{\circ}$. 
This sharp transition suggests a change in the dominant mode excitement mechanism as the polarization deviates from the principal axis of the nanostructure. 
This behavior demonstrates that the trapped-mode resonance maintains considerable stability against variations in the incident beam's polarization, suggesting potential applications in polarization-tolerant nonlinear optical devices.
The saturation value of $n_{2{\mathrm{eff}}}/n_{2{\mathrm{bulk}}}\approx 60$ at larger angles still represents significant enhancement compared to bulk GaP, indicating that some nonlinear enhancement mechanism remains active even when the polarization is misaligned with the structure principal axis. 
This residual enhancement may be due to partial excitation of the trapped state or contributions from other resonant modes of the nanostructure.
Figures~\ref{TRspectr_GaP_rotated_srot10}--\ref{efield_dist_GaP_144_lattrecthol_srot60} (Supplement~1) illustrate the transmittance and reflectance spectra, multipole decomposition results, time-averaged electric ($\varepsilon|\mathbf{E}|^2$) and magnetic ($|\mathbf{H}|^2$) energy distributions, and electric field distributions at the cross-section of the metasurface ($h/2$) for the polarization angles of $10^{\circ}$, $20^{\circ}$, $30^{\circ}$, and $60^{\circ}$.
The electric energy distribution pattern persists up to polarization angles of approximately $20^{\circ}$ (Fig.~\ref{ener_dist_GaP_144_lattrecthol_rot20}, Supplement~1), beyond which both the pattern fidelity and maximum energy density degrade (Fig.~\ref{efield_dist_GaP_144_lattrecthol_srot60}, Supplement~1).
The sharp transition observed at $17^\circ$ suggests potential applications in polarization-sensitive switching devices, where small changes in input polarization could induce large changes in nonlinear response.

\begin{figure}[tb!]
{\centering
\includegraphics[width=\subfigsize]{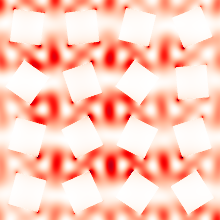}\hspace{1em}
\includegraphics[width=\subfigsize]{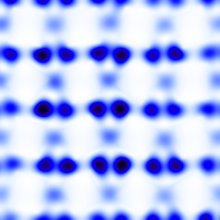}\par}
{\centering
\includegraphics[width=\subfigsize]{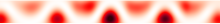}\hspace{1em}
\includegraphics[width=\subfigsize]{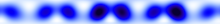}\par}
\caption{\label{ener_dist_GaP_144_lattrecthol_ra17.5} 
Time-averaged distributions of the electric ($\varepsilon|\mathbf{E}|^2$; left panel, red) and magnetic ($|\mathbf{H}|^2$; right panel, blue) energy densities in the lattice of rotated rectangular nanoholes in a GaP slab with $a = 250$~nm, $h = 100$~nm, $b_1 = 140$~nm, and $b_2 = 148$~nm at $\lambda = 532$~nm. 
Each nanohole has a base rotation of $\beta = 20^\circ$ with an additional random rotation up to $\pm 17.5^\circ$ applied. 
Cross-sectional views show the distributions at $z = h/2$ in the $y$-$x$ plane (upper row, under vertically polarized Gaussian beam illumination) and at $x=0$ in the $y$-$z$ plane (lower row).
The colormap is same as that used in Fig.~\ref{ener_dist_GaP_144_lattrecthol}.}
\end{figure}

Next, the stability  of the nonlinear optical properties of the metasurface under study to geometric irregularities is investigated.
These irregularities are modeled by introducing random rotations of each nanohole about its center, with additional angles applied in both clockwise and counterclockwise directions. 
The rotation magnitude is constrained by a specified maximum angle.
Figure~
\ref{n2_yrot_GaP_144_lattrecthol}(b)
presents the simulation results for this analysis.
The data reveal that the enhancement ratio $n_{2\mathrm{eff}}/n_{2\mathrm{bulk}}$ remains relatively stable for maximum random rotation angles up to $8^{\circ}$. 
Beyond this threshold, $n_{2\mathrm{eff}}$ exhibits a gradual decrease, indicating that the nonlinear optical properties are sensitive to geometric irregularities in the metasurface. 
In particular, this operational range encompasses typical fabrication tolerances for all-dielectric optical metasurfaces.
The corresponding energy distributions are shown in Fig.~\ref{ener_dist_GaP_144_lattrecthol_ra17.5}. 
The figure demonstrates that while some lattice elements maintain energy distributions similar to those in the regular lattice (albeit with reduced magnitude), others show significant deviations. 
Only a fraction of the meta-atoms retain the regular lattice's energy distribution pattern, resulting in an attenuated net effect.
When the nanohole edges acquire random orientations, the effective OKE not only diminishes but can even undergo sign reversal in some cases.
This transition to negative values of $n_{2\mathrm{eff}}$ may occur when most meta-atoms enter the regime where $n_{2\mathrm{eff}}$ displays a dip, as shown in Fig.~\ref{n2_hl_GaP_lattrecthol}(a). For random nanohole orientations, the sign of \( n_{2\mathrm{eff}} \) becomes chaotic.


Thus, the proposed metasurface with a non-BIC trapped-mode resonance exhibits stability against fabrication imperfections. 
The low quality factor (evidenced by the wide transmittance dip in Fig.~\ref{TRspectr_GaP_rotated}) does not limit the effective second-order refractive index enhancement. 
In contrast, metasurfaces with BIC, despite their theoretically predicted high quality factors, demonstrate only about two orders of magnitude enhancement in second harmonic generation compared to unpatterned samples in real experiments \cite{Liu19,Zhang22}. This limitation stems from fabrication imperfections, which are particularly critical for large-area quasi-BIC samples \cite{Maslova21}.
Due to its strong reflection properties, the proposed nanostructure is suitable for reflective applications, such as high harmonic generation. These nonlinear optical phenomena, as well as OKE, depend on the enhancement of electric energy density within the metasurface.


To summarize, this work demonstrates that a GaP metasurface with a non-BIC trapped-mode resonance exhibits remarkable enhancement of the optical Kerr effect, achieving an effective second-order refractive index up to 700 times greater than bulk GaP. Comprehensive three-dimensional FDTD simulations reveal that this enhancement originates from strong electromagnetic field localization in rotated rectangular nanoholes arranged in a square lattice within a GaP slab.

The trapped-mode resonance displays a wide transmittance dip (low Q-factor) that remains robust against fabrication imperfections, in contrast to BIC-based metasurfaces where performance degrades significantly with structural disorder. These findings establish that low-Q trapped modes in dielectric metasurfaces can provide superior nonlinear enhancement compared to high-Q quasi-BIC resonances, particularly when considering fabrication imperfections.

Future work could explore optimization for other spectral ranges and materials with high intrinsic nonlinearities, as well as experimental realization of these structures.

\paragraph{Acknowledgement.} {\footnotesize
The research was carried out within the state assignment of IACP FEB RAS (Theme FWFW-2021-0001).
The results were obtained with the use of IACP FEB RAS Shared Resource Center ``Far Eastern Computing Resource'' equipment (https://www.cc.dvo.ru).}

\paragraph{Disclosures.} {\footnotesize
The author declares no conflicts of interest.}

\paragraph{Supplemental document.} {\footnotesize
See Supplement 1 for supporting content. }

\paragraph{Data availability.} {\footnotesize
Data underlying the results presented in this paper are not publicly available at this time but may be obtained from the authors upon reasonable request.}



\bibliography{nlphase}

\bibliographyfullrefs{nlphase}


\clearpage
\newpage
\setcounter{page}{1}
\setcounter{figure}{0}
\setcounter{equation}{0}
\renewcommand{\thefigure}{S\arabic{figure}}
\renewcommand{\thepage}{S\arabic{page}}

\setlength{\figsize}{7.5cm}
\setlength{\subfigsize}{3cm}



{\bfseries
\noindent
This document provides supplementary information for ``Enhanced Optical Kerr Effect in Metasurfaces Featuring Arrays of Rotated Rectangular Holes via Trapped-Mode Resonances.'' 
Additional graphs for the electric field and electromagnetic energy distributions, transmittance, reflectivity, and multipole decomposition results are shown.

 }


\begin{figure}[tbh!]
\begin{center}
{\centering
\includegraphics[width=\figsize]{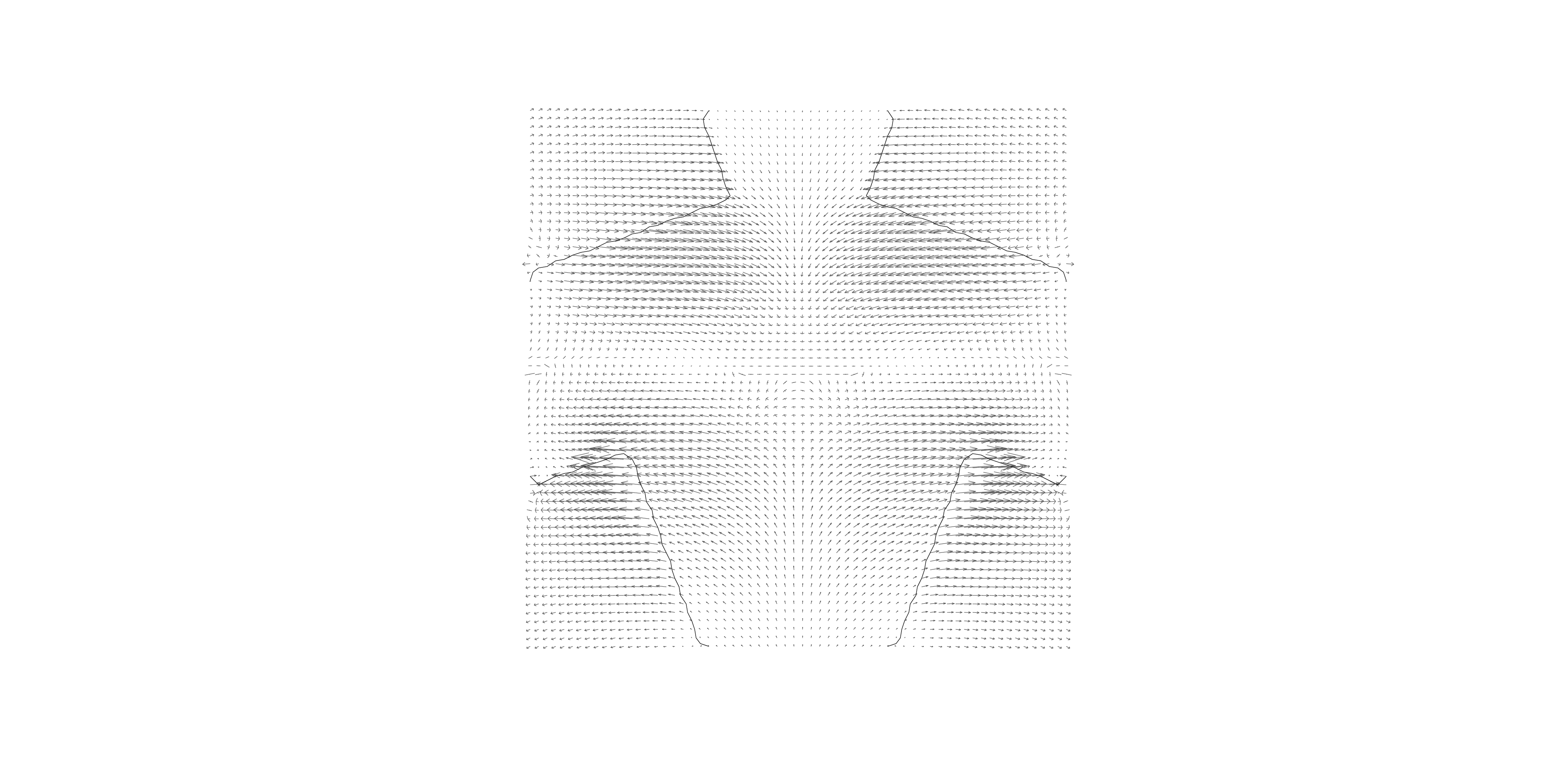}
\par
} 
\end{center}
\caption{\label{efield_dist_GaP_144_lattrecthol} 
Distribution of electric field for the rotated nanohole lattice element in the GaP slab (outlined in blue in Fig.~\ref{rect_doubang_pores}) lattice constant $a = 250$~nm, slab thickness $h = 100$~nm, nanohole dimensions $b_1 = 140$~nm and $b_2 = 148$~nm, rotation angle $\beta = 20^\circ$, and wavelength $\lambda = 532$~nm within the plane of $h/2$. The incident Gaussian beam is polarized vertically.
}
\end{figure} 

\begin{figure}[tb!]
{\centering
\includegraphics[width=\figsize]{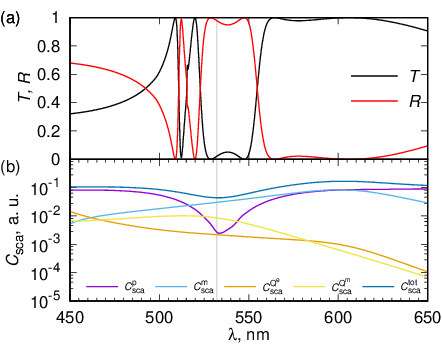}
\par} 
\caption{\label{TRspectr_GaP_rotated_bz1.38}
(a) Transmittance and reflectance spectra for the array of the rotated rectangular nanoholes. 
(b) Scattering cross-section spectra showing multipole contributions: electric dipole ($C_{\mathrm{sca}}^{\mathrm{p}}$), magnetic dipole ($C_{\mathrm{sca}}^{\mathrm{m}}$), electric quadrupole ($C_{\mathrm{sca}}^{\mathrm{Q^e}}$), and magnetic quadrupole ($C_{\mathrm{sca}}^{\mathrm{Q^m}}$), along with their sum ($C_{\mathrm{sca}}^{\mathrm{tot}}$). 
Results are shown for the rotated nanohole lattice element in the GaP slab (outlined in blue in Fig.~\ref{rect_doubang_pores})
Geometric parameters for both graphs: $a = 250$~nm, $h = 100$~nm, $b_1 = 140$~nm, $b_2 = 138$~nm, and $\beta = 20^\circ$. 
Refractive index $n_{0\,\mathrm{in}}$ is assumed to be constant over the whole wavelength range.}
\end{figure} 

\begin{figure}[tb!]
{\centering
\includegraphics[width=\figsize]{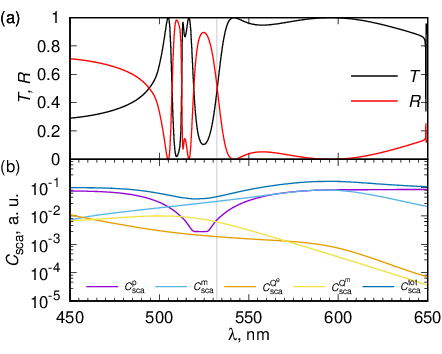}
\par} 
\caption{\label{TRspectr_GaP_rotated_bz1.54}
(a) Transmittance and reflectance spectra for the array of the rotated rectangular nanoholes. 
(b) Scattering cross-section spectra showing multipole contributions: electric dipole ($C_{\mathrm{sca}}^{\mathrm{p}}$), magnetic dipole ($C_{\mathrm{sca}}^{\mathrm{m}}$), electric quadrupole ($C_{\mathrm{sca}}^{\mathrm{Q^e}}$), and magnetic quadrupole ($C_{\mathrm{sca}}^{\mathrm{Q^m}}$), along with their sum ($C_{\mathrm{sca}}^{\mathrm{tot}}$). 
Results are shown for the rotated nanohole lattice element in the GaP slab (outlined in blue in Fig.~\ref{rect_doubang_pores})
Geometric parameters for both graphs: $a = 250$~nm, $h = 100$~nm, $b_1 = 140$~nm, $b_2 = 154$~nm, and $\beta = 20^\circ$. 
Refractive index $n_{0\,\mathrm{in}}$ is assumed to be constant over the whole wavelength range.}
\end{figure} 

\begin{figure}[tb!]
{\centering
\includegraphics[width=\figsize]{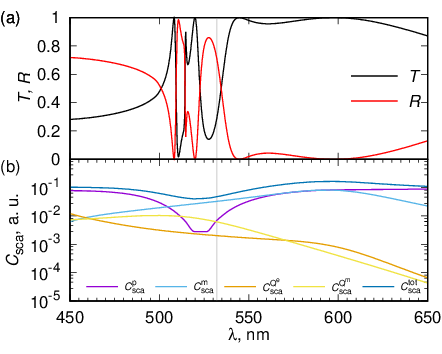}
\par} 
\caption{\label{TRspectr_GaP_rotated_da15}
(a) Transmittance and reflectance spectra for the array of the rotated rectangular nanoholes. 
(b) Scattering cross-section spectra showing multipole contributions: electric dipole ($C_{\mathrm{sca}}^{\mathrm{p}}$), magnetic dipole ($C_{\mathrm{sca}}^{\mathrm{m}}$), electric quadrupole ($C_{\mathrm{sca}}^{\mathrm{Q^e}}$), and magnetic quadrupole ($C_{\mathrm{sca}}^{\mathrm{Q^m}}$), along with their sum ($C_{\mathrm{sca}}^{\mathrm{tot}}$). 
Results are shown for the rotated nanohole lattice element in the GaP slab (outlined in blue in Fig.~\ref{rect_doubang_pores})
Geometric parameters for both graphs: $a = 250$~nm, $h = 100$~nm, $b_1 = 140$~nm, $b_2 = 148$~nm, and $\beta = 15^\circ$. 
Refractive index $n_{0\,\mathrm{in}}$ is assumed to be constant over the whole wavelength range.}
\end{figure} 

\begin{figure}[tb!]
{\centering
\includegraphics[width=\figsize]{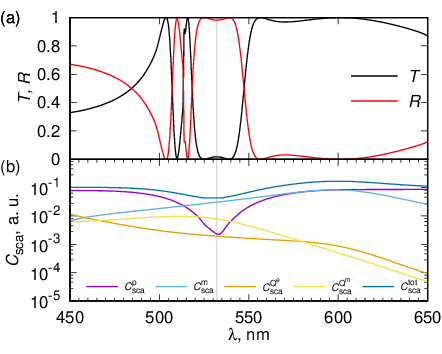}
\par} 
\caption{\label{TRspectr_GaP_rotated_da25}
(a) Transmittance and reflectance spectra for the array of the rotated rectangular nanoholes. 
(b) Scattering cross-section spectra showing multipole contributions: electric dipole ($C_{\mathrm{sca}}^{\mathrm{p}}$), magnetic dipole ($C_{\mathrm{sca}}^{\mathrm{m}}$), electric quadrupole ($C_{\mathrm{sca}}^{\mathrm{Q^e}}$), and magnetic quadrupole ($C_{\mathrm{sca}}^{\mathrm{Q^m}}$), along with their sum ($C_{\mathrm{sca}}^{\mathrm{tot}}$). 
Results are shown for the rotated nanohole lattice element in the GaP slab (outlined in blue in Fig.~\ref{rect_doubang_pores})
Geometric parameters for both graphs: $a = 250$~nm, $h = 100$~nm, $b_1 = 140$~nm, $b_2 = 148$~nm, and $\beta = 25^\circ$. 
Refractive index $n_{0\,\mathrm{in}}$ is assumed to be constant over the whole wavelength range.}
\end{figure} 

\begin{figure}[tb!]
{\centering
\includegraphics[width=\figsize]{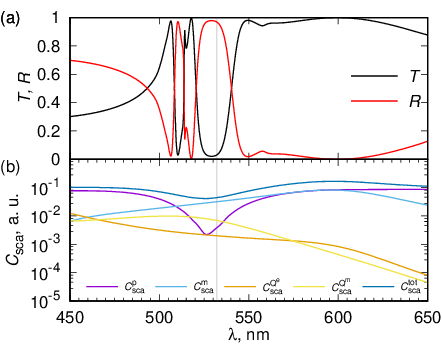}
\par} 
\caption{\label{TRspectr_GaP_rotated_srot10}
(a) Transmittance and reflectance spectra for the array of the rotated rectangular nanoholes. 
(b) Scattering cross-section spectra showing multipole contributions: electric dipole ($C_{\mathrm{sca}}^{\mathrm{p}}$), magnetic dipole ($C_{\mathrm{sca}}^{\mathrm{m}}$), electric quadrupole ($C_{\mathrm{sca}}^{\mathrm{Q^e}}$), and magnetic quadrupole ($C_{\mathrm{sca}}^{\mathrm{Q^m}}$), along with their sum ($C_{\mathrm{sca}}^{\mathrm{tot}}$). 
Results are shown for the rotated nanohole lattice element in the GaP slab (outlined in blue in Fig.~\ref{rect_doubang_pores})
Geometric parameters for both graphs: $a = 250$~nm, $h = 100$~nm, $b_1 = 140$~nm, $b_2 = 148$~nm, and $\beta = 20^\circ$. 
The incident beam is polarized at $10^\circ$ with respect to the vertical axis.
Refractive index $n_{0\,\mathrm{in}}$ is assumed to be constant over the whole wavelength range.}
\end{figure} 

\begin{figure}[tb!]
{\centering
\includegraphics[width=\subfigsize]{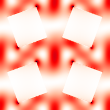}\hspace{1em}
\includegraphics[width=\subfigsize]{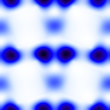}\par}
\caption{\label{ener_dist_GaP_144_lattrecthol_rot10} 
Time-averaged distributions of the electric ($\varepsilon|\mathbf{E}|^2$; left panel, red) and magnetic ($|\mathbf{H}|^2$; right panel, blue) energy densities in the lattice of rotated rectangular nanoholes in the GaP slab with $a = 250$~nm, $h = 100$~nm, $b_1 = 140$~nm, $b_2 = 148$~nm,  $\beta = 20^\circ$ at $\lambda = 532$~nm. 
The distributions are calculated in the cross-sectional plane at $h/2$. 
The incident Gaussian beam is polarized at $10^\circ$ with respect to the vertical axis.
The colormap is same as that used in Fig.~\ref{ener_dist_GaP_144_lattrecthol}.}
\end{figure} 

\begin{figure}[tbh!]
\begin{center}
{\centering
\includegraphics[width=\figsize]{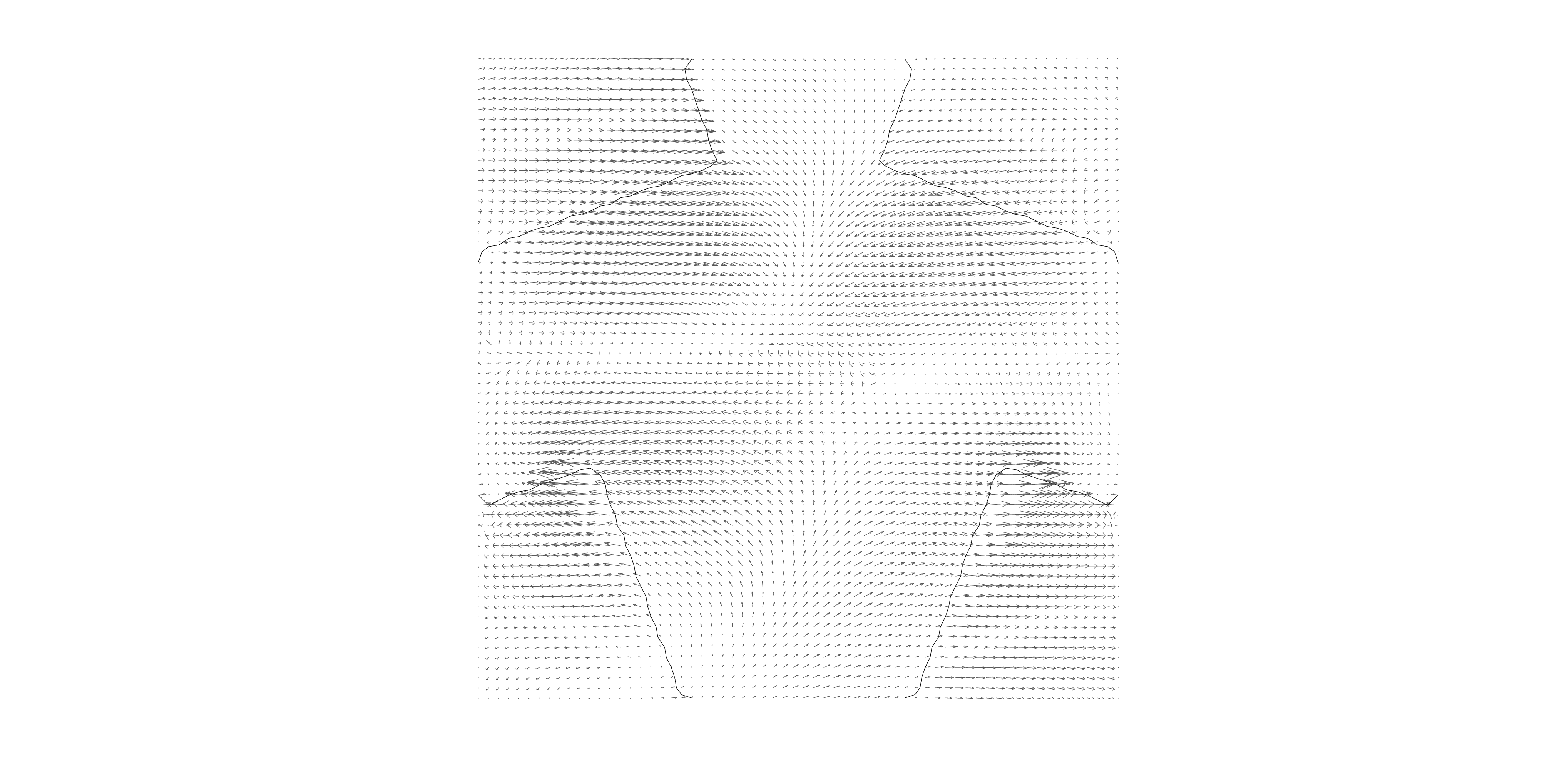}
\par
} 
\end{center}
\caption{\label{efield_dist_GaP_144_lattrecthol_srot10} 
Distribution of electric field for the rotated nanohole lattice element in the GaP slab (outlined in blue in Fig.~1) lattice constant $a = 250$~nm, slab thickness $h = 100$~nm, nanohole dimensions $b_1 = 140$~nm and $b_2 = 148$~nm, rotation angle $\beta = 20^\circ$, and wavelength $\lambda = 532$~nm within the plane of $h/2$. The incident Gaussian beam is polarized at $10^\circ$ with respect to the vertical axis.
}
\end{figure} 

\begin{figure}[tb!]
{\centering
\includegraphics[width=\figsize]{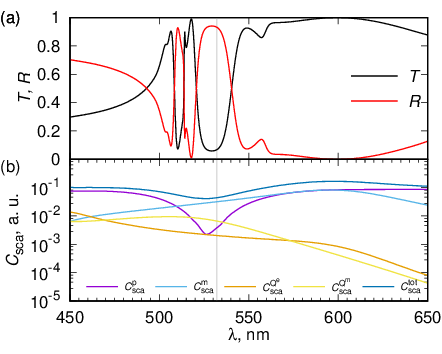}
\par} 
\caption{\label{TRspectr_GaP_rotated_srot20}
(a) Transmittance and reflectance spectra for the array of the rotated rectangular nanoholes. 
(b) Scattering cross-section spectra showing multipole contributions: electric dipole ($C_{\mathrm{sca}}^{\mathrm{p}}$), magnetic dipole ($C_{\mathrm{sca}}^{\mathrm{m}}$), electric quadrupole ($C_{\mathrm{sca}}^{\mathrm{Q^e}}$), and magnetic quadrupole ($C_{\mathrm{sca}}^{\mathrm{Q^m}}$), along with their sum ($C_{\mathrm{sca}}^{\mathrm{tot}}$). 
Results are shown for the rotated nanohole lattice element in the GaP slab (outlined in blue in Fig.~\ref{rect_doubang_pores})
Geometric parameters for both graphs: $a = 250$~nm, $h = 100$~nm, $b_1 = 140$~nm, $b_2 = 148$~nm, and $\beta = 20^\circ$. 
The incident beam is polarized at $20^\circ$ with respect to the vertical axis.
Refractive index $n_{0\,\mathrm{in}}$ is assumed to be constant over the whole wavelength range.}
\end{figure} 

\begin{figure}[tb!]
{\centering
\includegraphics[width=\subfigsize]{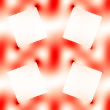}\hspace{1em}
\includegraphics[width=\subfigsize]{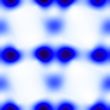}\par}
\caption{\label{ener_dist_GaP_144_lattrecthol_rot20} 
Time-averaged distributions of the electric ($\varepsilon|\mathbf{E}|^2$; left panel, red) and magnetic ($|\mathbf{H}|^2$; right panel, blue) energy densities in the lattice of rotated rectangular nanoholes in the GaP slab with $a = 250$~nm, $h = 100$~nm, $b_1 = 140$~nm, $b_2 = 148$~nm,  $\beta = 20^\circ$ at $\lambda = 532$~nm. 
The distributions are calculated in the cross-sectional plane at $h/2$. 
The incident Gaussian beam is polarized at $20^\circ$ with respect to the vertical axis.
The colormap is same as that used in Fig.~\ref{ener_dist_GaP_144_lattrecthol}.}
\end{figure} 

\begin{figure}[tbh!]
\begin{center}
{\centering
\includegraphics[width=\figsize]{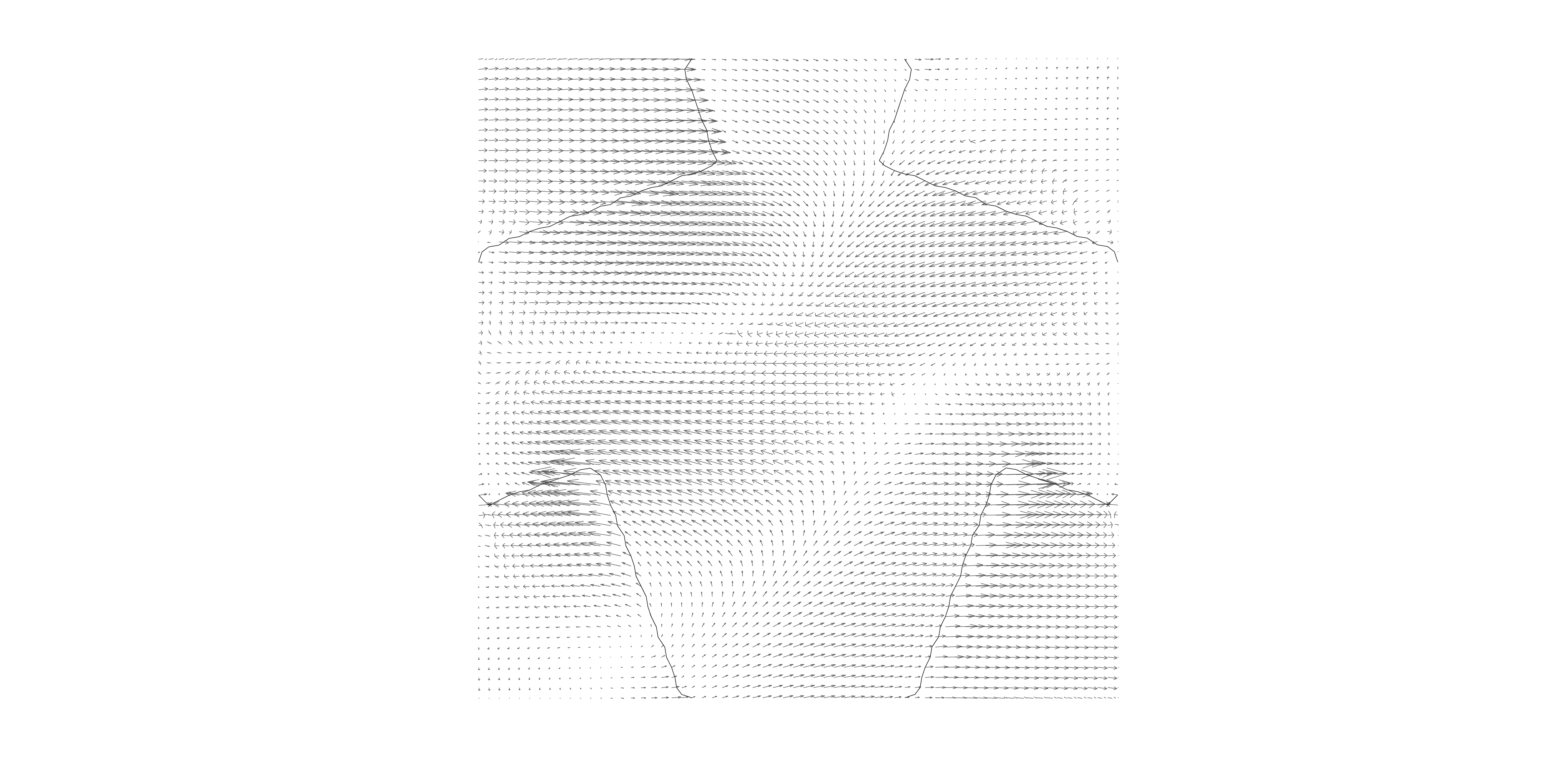}
\par
} 
\end{center}
\caption{\label{efield_dist_GaP_144_lattrecthol_srot20} 
Distribution of electric field for the rotated nanohole lattice element in the GaP slab (outlined in blue in Fig.~1) lattice constant $a = 250$~nm, slab thickness $h = 100$~nm, nanohole dimensions $b_1 = 140$~nm and $b_2 = 148$~nm, rotation angle $\beta = 20^\circ$, and wavelength $\lambda = 532$~nm within the plane of $h/2$. The incident Gaussian beam is polarized at $20^\circ$ with respect to the vertical axis.
}
\end{figure} 

\begin{figure}[tb!]
{\centering
\includegraphics[width=\figsize]{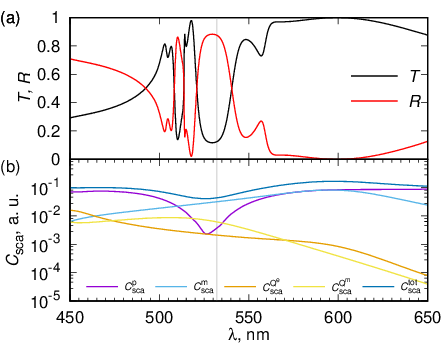}
\par} 
\caption{\label{TRspectr_GaP_rotated_srot30}
(a) Transmittance and reflectance spectra for the array of the rotated rectangular nanoholes. 
(b) Scattering cross-section spectra showing multipole contributions: electric dipole ($C_{\mathrm{sca}}^{\mathrm{p}}$), magnetic dipole ($C_{\mathrm{sca}}^{\mathrm{m}}$), electric quadrupole ($C_{\mathrm{sca}}^{\mathrm{Q^e}}$), and magnetic quadrupole ($C_{\mathrm{sca}}^{\mathrm{Q^m}}$), along with their sum ($C_{\mathrm{sca}}^{\mathrm{tot}}$). 
Results are shown for the rotated nanohole lattice element in the GaP slab (outlined in blue in Fig.~\ref{rect_doubang_pores})
Geometric parameters for both graphs: $a = 250$~nm, $h = 100$~nm, $b_1 = 140$~nm, $b_2 = 148$~nm, and $\beta = 60^\circ$. 
The incident beam is polarized at $30^\circ$ with respect to the vertical axis.
Refractive index $n_{0\,\mathrm{in}}$ is assumed to be constant over the whole wavelength range.}
\end{figure} 

\begin{figure}[tb!]
{\centering
\includegraphics[width=\subfigsize]{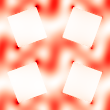}\hspace{1em}
\includegraphics[width=\subfigsize]{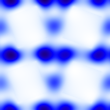}\par}
\caption{\label{ener_dist_GaP_144_lattrecthol_rot30} 
Time-averaged distributions of the electric ($\varepsilon|\mathbf{E}|^2$; left panel, red) and magnetic ($|\mathbf{H}|^2$; right panel, blue) energy densities in the lattice of rotated rectangular nanoholes in the GaP slab with $a = 250$~nm, $h = 100$~nm, $b_1 = 140$~nm, $b_2 = 148$~nm,  $\beta = 20^\circ$ at $\lambda = 532$~nm. 
The distributions are calculated in the cross-sectional plane at $h/2$. 
The incident Gaussian beam is polarized at $30^\circ$ with respect to the vertical axis.
The colormap is same as that used in Fig.~\ref{ener_dist_GaP_144_lattrecthol}.}
\end{figure} 

\begin{figure}[tbh!]
\begin{center}
{\centering
\includegraphics[width=\figsize]{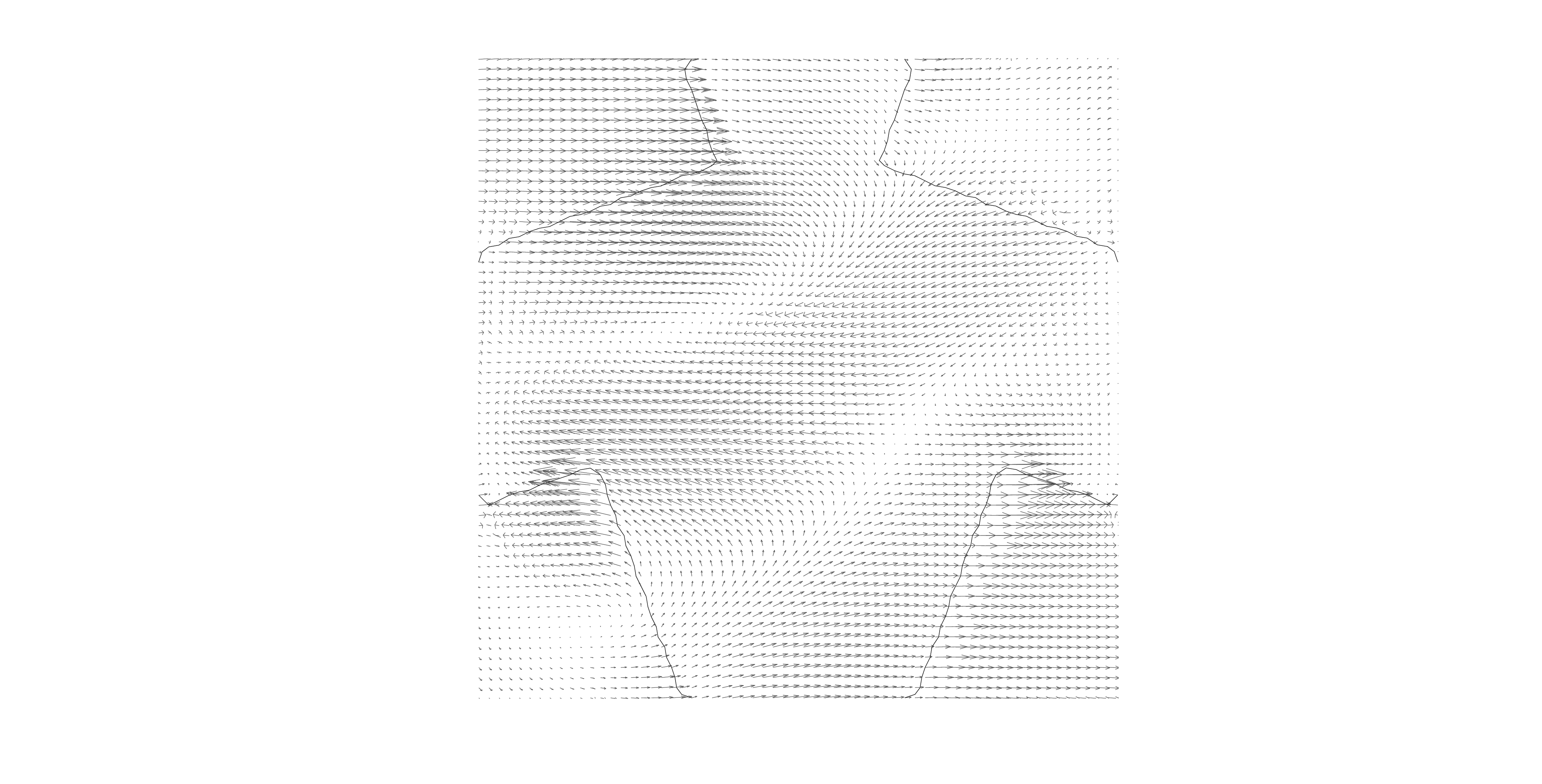}
\par
} 
\end{center}
\caption{\label{efield_dist_GaP_144_lattrecthol_srot30} 
Distribution of electric field for the rotated nanohole lattice element in the GaP slab (outlined in blue in Fig.~1) lattice constant $a = 250$~nm, slab thickness $h = 100$~nm, nanohole dimensions $b_1 = 140$~nm and $b_2 = 148$~nm, rotation angle $\beta = 20^\circ$, and wavelength $\lambda = 532$~nm within the plane of $h/2$. The incident Gaussian beam is polarized at $30^\circ$ with respect to the vertical axis.
}
\end{figure} 

\begin{figure}[tb!]
{\centering
\includegraphics[width=\figsize]{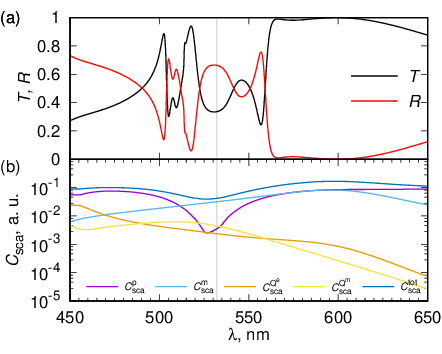}
\par} 
\caption{\label{TRspectr_GaP_rotated_srot60}
(a) Transmittance and reflectance spectra for the array of the rotated rectangular nanoholes. 
(b) Scattering cross-section spectra showing multipole contributions: electric dipole ($C_{\mathrm{sca}}^{\mathrm{p}}$), magnetic dipole ($C_{\mathrm{sca}}^{\mathrm{m}}$), electric quadrupole ($C_{\mathrm{sca}}^{\mathrm{Q^e}}$), and magnetic quadrupole ($C_{\mathrm{sca}}^{\mathrm{Q^m}}$), along with their sum ($C_{\mathrm{sca}}^{\mathrm{tot}}$). 
Results are shown for the rotated nanohole lattice element in the GaP slab (outlined in blue in Fig.~\ref{rect_doubang_pores})
Geometric parameters for both graphs: $a = 250$~nm, $h = 100$~nm, $b_1 = 140$~nm, $b_2 = 148$~nm, and $\beta = 60^\circ$. 
The incident beam is polarized at $20^\circ$ with respect to the vertical axis.
Refractive index $n_{0\,\mathrm{in}}$ is assumed to be constant over the whole wavelength range.}
\end{figure} 

\begin{figure}[tb!]
{\centering
\includegraphics[width=\subfigsize]{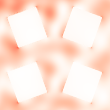}\hspace{1em}
\includegraphics[width=\subfigsize]{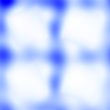}\par}
\caption{\label{ener_dist_GaP_144_lattrecthol_rot60} 
Time-averaged distributions of the electric ($\varepsilon|\mathbf{E}|^2$; left panel, red) and magnetic ($|\mathbf{H}|^2$; right panel, blue) energy densities in the lattice of rotated rectangular nanoholes in the GaP slab with $a = 250$~nm, $h = 100$~nm, $b_1 = 140$~nm, $b_2 = 148$~nm,  $\beta = 20^\circ$ at $\lambda = 532$~nm. 
The distributions are calculated in the cross-sectional plane at $h/2$. 
The incident Gaussian beam is polarized at $60^\circ$ with respect to the vertical axis.
The colormap is same as that used in Fig.~\ref{ener_dist_GaP_144_lattrecthol}.}
\end{figure} 

\begin{figure}[tbh!]
\begin{center}
{\centering
\includegraphics[width=\figsize]{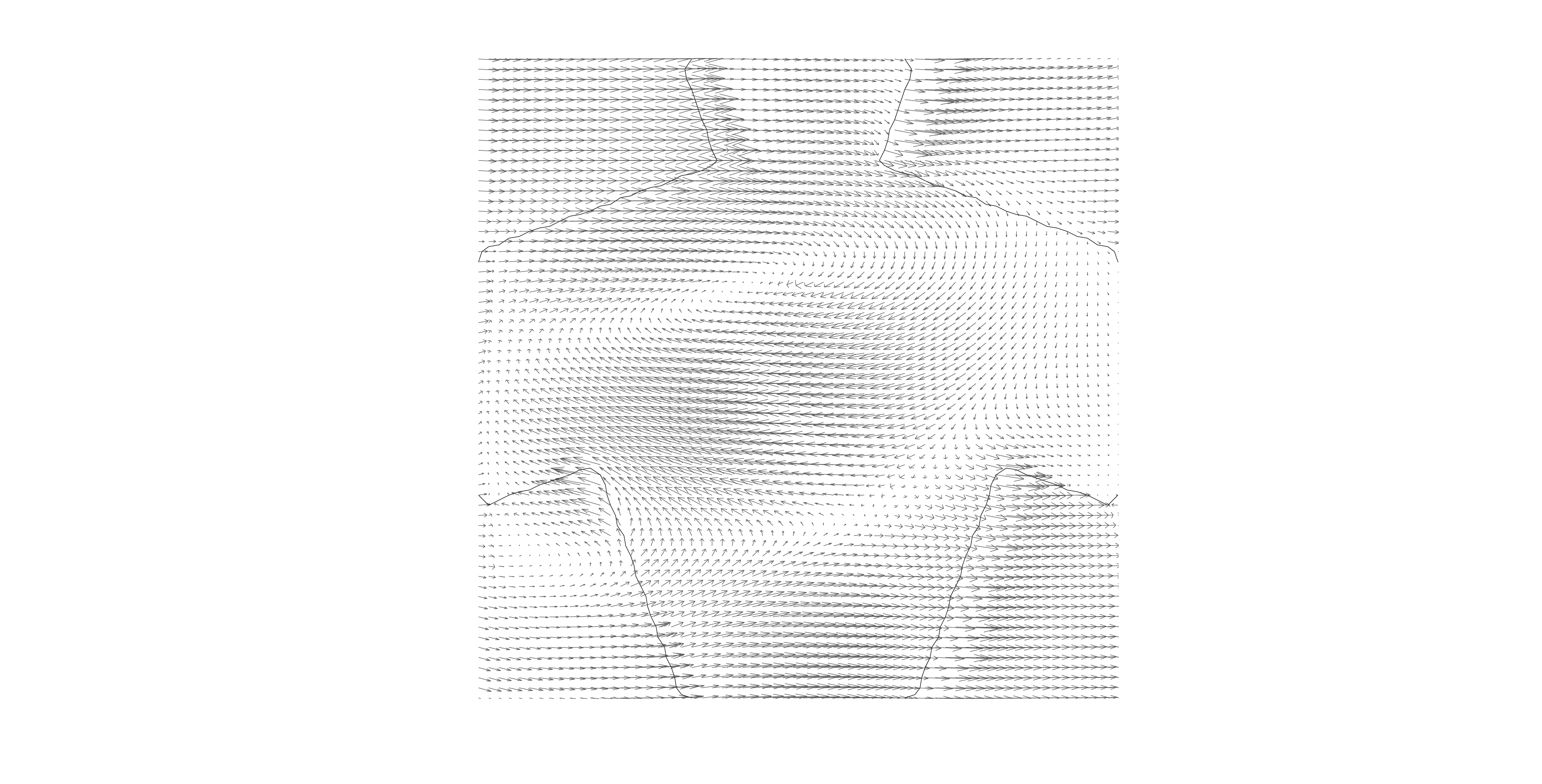}
\par
} 
\end{center}
\caption{\label{efield_dist_GaP_144_lattrecthol_srot60} 
Distribution of electric field for the rotated nanohole lattice element in the GaP slab (outlined in blue in Fig.~\ref{rect_doubang_pores}) lattice constant $a = 250$~nm, slab thickness $h = 100$~nm, nanohole dimensions $b_1 = 140$~nm and $b_2 = 148$~nm, rotation angle $\beta = 20^\circ$, and wavelength $\lambda = 532$~nm within the plane of $h/2$. The incident Gaussian beam is polarized at $60^\circ$ with respect to the vertical axis.
}
\end{figure}



\end{document}